\documentclass[prd,aps,floats,twocolumn,nofootinbib]{revtex4-1}
\usepackage{slashed}
\usepackage{mathtools}
\usepackage{amsfonts}
\usepackage{amssymb}
\usepackage{epsfig}
\usepackage{rotating}
\usepackage{url}
\usepackage{times}
\usepackage{color}
\usepackage{bm}
\usepackage{xcolor,colortbl}
\usepackage{hyperref}
\usepackage[alsoload=hep]{siunitx}
\usepackage{enumitem}
\usepackage{fancyhdr}
\usepackage{anyfontsize}
\usepackage{wasysym}
\usepackage{empheq}
\usepackage{float}
\usepackage{bm}
\newcommand{\plk}{\mathfrak{h}}
\newcommand{\be}{\begin{equation}}
\newcommand{\ee}{\end{equation}}
\newcommand{\bea}{\begin{eqnarray}}
\newcommand{\eea}{\end{eqnarray}}
\newcommand{\nn}{\nonumber}
\newcommand{\fn}{\footnote}

\begin{document}

\title{Semiclassical limit problems with concurrent use of several clocks in quantum cosmology}
\author{Bruno Alexandre}
\author{Jo\~{a}o Magueijo}
\email{magueijo@ic.ac.uk}
\affiliation{Theoretical Physics Group, The Blackett Laboratory, Imperial College, Prince Consort Rd., London, SW7 2BZ, United Kingdom}

\date{\today}

\begin{abstract}
We revisit a recent proposal for a definition of time in quantum cosmology, to investigate the effects of 
having more than one possible type of clock ``at the same time''. We use as test tube an extension of Einstein gravity with a massless scalar field in which the gravitational coupling $G_N$  is only a constant on-shell, mimicking the procedure for $\Lambda$ in unimodular gravity. 
Hence we have two ``simultaneous'' clocks in the theory: a scalar field clock, and the conjugate of $G_N$.  We find that
attempts to use two coherent clocks concurrently are disastrous for recovering the classical limit. The Heisenberg relations, instead of being
saturated, are always realized abundantly above their bound, with strong quantum effects expected at least in parts of the trajectory. Semi-classical 
states always result from situations where we effectively impose a single clock, either by making the other clock a failed clock (i.e. by choosing
a state where its conjugate constant is infinitely sharp), or by choosing a basis of constants where all clocks but one are redundant, i.e. motion or change in  phase space does not occur with the passing of their ``times''. We show how this conclusion generalizes to fluids with any equation of state. It also applies to systems where ``sub-clocks'' of the same type could be  used, for example in mixtures of different fluids with the same equation of state. 
\end{abstract}

\maketitle

\section{Introduction}
The idea that there might be more than one time strand running in parallel is either anathema (in physics) or self-evident (in human experience, in literature, etc). Yet, the conflict between general relativity and quantum mechanics, in its implications for the meaning of time \cite{Isham,Kuch,selftaught,marina,rovelli,ebb}, has created a situation where the universe is either timeless or plagued by the appearance of several relational physical times. This is far from innocuous, e.g. for the discussion of whether there is a physical primordial singularity in our Universe~\cite{gielen,gielen1}. 

In recent work~\cite{Jletter} it was proposed that the constants of Nature might provide excellent relational quantum clocks in the form of their conjugate momenta in theories where they are only constant on-shell. In this context the ubiquitous embarrassment of riches follows from the plethora of constants \cite{Barrowtip} that can act as progenitors of time. In~\cite{Jletter,JPRD} this was embraced, rather than regretted, following the view that different clocks should be used at different stages in the life of the Universe, with the adjustment of clocks across ``time zones'' to be seen as a feature of our Universe. ``Minority'' clocks always exist but should be abandoned to give way to the dominant clocks. Arguments related to unitary evolution and preservation of a semi-classical limit were provided in favour of this view.

In~\cite{JPRD} we did not consider the possibility that several dominant clocks might concur in the same epoch. What would be the implications?
In this paper we start to answer this question availing ourselves of a few simple toy models. Our main finding is that the more clocks we use concurrently, the more difficult it becomes to find a semi-classical limit. We examine the Heisenberg uncertainty bound, usually saturated by coherent states, and show how in all but one case balanced saturation of the bound becomes elusive once several dominant clocks are at play.

The plan of this paper is as follows. In Section~\ref{simplemodel} we start by presenting our simplest illustrative model, where the gravitational coupling $G_N$ is deconstantized (following a procedure similar to unimodular gravity and the deconstantization of $\Lambda$). Thus, the momentum of $G_N$ provides a dominant clock in addition to that supplied by the massless scalar field. We then subject the model to the treatment proposed in~\cite{Jletter,JPRD}, identifying the `two-time'' Schrodinger equation, and the most general wave-functions (Section~\ref{quant}). These are superpositions of mono-chromatic partial waves in two frequencies and times, with amplitudes in the two ``de-constants'' ($G_N$ and the momentum $\pi_\phi$ of the field). 

The core of our paper is in Section~\ref{twotime}, where we examine how concrete solutions interact with the Heisenberg uncertainty bound and its possible saturation. We can of course consider an infinitely sharp wave function in one constant (so that its conjugate clock becomes uniformly distributed, and so it is a failed clock), but this effectively reduces the two-clock system to a single one. Using two coherent states results in disaster, with a surplus uncertainty over the bound for each clock, which blows up in some regions of connection space. Any attempts to bypass this disaster always reduce the two-clock system to a single one (for example considering coherent states in the frequency conjugate to the wave number; see Section~\ref{field-redef} for details). One clock is better than two. 

The rest of the paper is spent generalizing this result and encountering it in different settings. In Section~\ref{genfluids}, we consider scalar fields with non-quadratic kinetic terms (i.e. $K$-essence~\cite{k-ess}), with equations of state $w$ other than $w=1$. The dependence on $G_N$ is the same as in other perfect fluid models (e.g.~\cite{brown,gielen,gielen1}). In Section~\ref{multicomp} we consider a mixture of fluids with the same $w$, so that each can provide a dominating clock. The same difficulties in finding coherent states are found in all cases, with the exception of a multi-radiation fluid, where the wave function miraculously factors into semi-classical states in each of the sub-clocks. 

In two concluding Sections we examine the relation between this paper and other approaches based on the WKB limit, and we digress on the implications of our result, in particular regarding the cosmological constant problem.

\section{A simple model}
\label{simplemodel}

It is well known that a massless scalar field can provide an excellent relational clock, dual to a constant conjugate momentum (e.g.~\cite{gielen,gielen1,dit1,dit2}). 
We apply the prescription in~\cite{Jletter} to a toy model with Einstein gravity and a massless scalar field, in which the gravitational coupling to matter
is deconstantized according to the prescription used for $\Lambda$ in unimodular gravity \cite{henne,padi,unruh,padilla,vikman,vikman1}. That is, we take as base action:
\begin{eqnarray}
S_0=\frac{1}{16\pi G_0}\int d^4x\sqrt{-g}(R-8\pi G_N\partial_\mu\phi\partial^\mu\phi), 
\label{actionphi}
\end{eqnarray}
where the Planck length (determined by $G_0$) is kept fixed, but the gravitational coupling to matter, determined by $G_N$ is subject to 
deconstantization according to:
\be\label{S}
S_0\rightarrow S=S_0-\int d^4x\,\alpha\partial_\mu T^\mu =
S_0+\int d^4x\, T^\mu\partial_\mu \alpha
\ee
where $\alpha_G=\alpha_G(G_N)$ is an arbitrary function of $G_N$ at this stage (and the last identity is valid up to a boundary term which we shall ignore).
Different choices of $\alpha_G(G_N)$ and associated $T$ are related by a canonical transformation, leading to a classically (but not quantum mechanically) equivalent theories.

We will examine this model in mini-superspace, so that:
%
\begin{eqnarray}
S_0&=& \frac{3V_c}{8\pi G_0}\int dt\left[\dot{b}a^2+\dot{\phi}\pi_\phi-Na\left(-(b^2+k)+\frac{\pi_\phi^2}{2 \tilde G_Na^4}\right)\right] \nn
\end{eqnarray}
where $k=0,\pm1$ is the normalized spatial curvature, 
$a$ is the expansion factor, the connection variable $b$ is
the off-shell version of the Hubble parameter (since $b\approx \dot a$ if there is no torsion), 
the Lagrange multiplier $N$ is the lapse function and $V_c=\int d^3 x$ is the comoving volume of the region 
under study (which could be the whole manifold, should this be compact). 
In addition:
\begin{eqnarray}
&& \pi_\phi=\tilde{G}_N\dot{\phi}\frac{a^3}{N} \label{piphi},  \\
&& \tilde{G}_N\equiv\frac{8\pi G_N}{3}. 
\end{eqnarray}
This implies the  Poisson brackets with normalization:
\be
\{b,a^{2}\}=\{\phi,\pi_\phi\}=\frac{8\pi G_0}{3 V_{c}}.
\ee
The  Hamiltonian constraint is:
\begin{eqnarray}
H&=&Na\left( -(b^2+k)+\frac{\pi_\phi^2}{2a^4\tilde{G}_N}\right)=0, 
\label{hamphi}
\end{eqnarray}
which can be placed in the form proposed in~\cite{Jletter}:
\begin{eqnarray}
&& h(b)a^2-\alpha_1\alpha_2=0
\label{constraint}
\end{eqnarray}
with:
\begin{eqnarray}
&& h(b)=\sqrt{b^2+k}\label{hofb} \\
&& \alpha_1=\frac{1}{\sqrt{2\tilde{G}_N}} \label{alpha1} \\
&& \alpha_2=\pi_\phi \label{alpha2}
\end{eqnarray}
corresponding to a fluid with $w=1$. The $\alpha(G_N)$ in (\ref{S}) should therefore be proportional to $\alpha_1$, with the proportionality constant adjusted so that in mini-superspace:
\begin{equation}
S_0\rightarrow S=S_0+\frac{3V_c}{8\pi G_0}\int dt\dot{\alpha_1}T_1
\end{equation}
leading to the new PB:
\be
\{\alpha_1,T_1\}=\frac{8\pi G_0}{3 V_{c}}.
\ee
Hence we have a theory with two relational times, $T_1$ which is a conjugate of a function of $G_N$, and $T_2=\phi$, conjugate to the 
on-shell constant $\pi_\phi$.

\section{Quantization and Solutions}
\label{quant}
Quantization can now proceed just as in \cite{JPRD}, with the Poisson brackets becoming commutators
\begin{equation} \label{com1}
\left[ \hat{b},\hat{a^{2}}\right] =
\left[ \hat{\alpha_1},\hat{T_1}\right] =
\left[ \hat{\alpha_2},\hat{T_2}\right] =
i\plk
\end{equation}%
where $\plk=l_P^2/3V_c$, with the reduced Planck $l_P$ determined by the fixed $G_0$. In the representation diagonalizing
$b$, $T_1$ and $T_2$ we therefore transform the Hamiltonian constraint, not in an timeless Wheeler-DeWitt equation, but in a 
double-time Schroedinger equation:
\be\label{doubleTScho}
\left(-i\plk h(b)\frac{\partial}{\partial b} +\plk^2 \frac{\partial^2}{\partial T_1 \partial T_2}\right)\psi=0.
\ee
This is only a minor extension of the results in \cite{JPRD}, the upshot of which is that the general wave function takes the form:
\begin{widetext}
\begin{eqnarray}
\psi(b;T_1,T_2)&=&\int d\alpha_1d\alpha_2A(\alpha_1,\alpha_2)e^{-\frac{i}{\plk}[\alpha_1T_1+\alpha_2T_2]}\psi_s(b,\alpha_1,\alpha_2)\nn \\
&=& \int \frac{d\alpha_1d\alpha_2}{2\pi\plk}A(\alpha_1,\alpha_2)e^{\frac{i}{\plk}[X(b)\alpha_1\alpha_2-\alpha_1T_1-\alpha_2T_2]},
\label{psi}
\end{eqnarray}
\end{widetext}
where the spatial $\psi_s$ is a suitably normalized solution to the timeless Wheeler-DeWitt equation:
\be\label{psis}
\psi_s(b,\alpha_1,\alpha_2)=\frac{e^{\frac{i}{\plk}\alpha_1\alpha_2 X(b)}}{2\pi\plk}
\ee
and is a plane wave in the linearizing variable~\cite{DSR}:
\begin{eqnarray}
X(b)=\int\frac{db}{h(b)} && = \text{arctanh}\frac{b}{\sqrt{b^2+k}} , \;\; k>0 \nn\\
&& = \log b , \;\; k=0 \nn\\
&& = \text{arctanh}\frac{b}{\sqrt{b^2-|k|}}  , \;\; k<0.
\label{Xofb}
\end{eqnarray}
associated with $w=1$. 

\section{A two-time Heisenberg uncertainty principle}
\label{twotime}
The Heisenberg uncertainty principle is a purely kinematical result involving any canonical pair, and as such we have for a two-time,
two-constant setting:
\be\label{Heis0}
\sigma_{T_i}\sigma_{\alpha_i} \ge  \frac{ \plk}{ 2 }.
\ee
The question, however, is how the dynamical solutions to the theory make these two sets of uncertainties interact:
Can we saturate them simultaneously? Or do multiple clocks get in each other's way?
Rather than providing a formal analysis we present a number of physically obvious solutions and examine how they relate to 
the Heisenberg uncertainity bounds.

\subsection{One failed and one coherent clock} 
As a warm up we consider the product of a coherent (semi-classical) state in one clock/constant, and an infinitely sharp constant (i.e. a  failed clock) in the other:
\begin{eqnarray}
&& A(\alpha_1,\alpha_2)=\delta(\alpha_1-\alpha_{10})\sqrt{\boldsymbol{N}(\alpha_{20},\sigma)} .
\end{eqnarray}
Unsurprisingly, integrating (\ref{psi}) leads to the usual coherent state in the second variable, 
i.e., the product of a plane wave centered on ${\bm \alpha}_0$ and a Gaussian envelope in the coherent variable:
\begin{eqnarray}
\psi(b,T_1,T_2)&=&Ne^{-\frac{\sigma^2}{\plk^2}(\alpha_{10}X-T_2)^2+\frac{i}{\plk}[\alpha_{20}\alpha_{10}X-\alpha_{10}T_1-\alpha_{20}T_2]}\nn\\
&=&\psi(b,T_1,T_2;{\bm \alpha}_0)e^{-\frac{\sigma^2}{\plk^2}(\alpha_{10}X-T_2)^2}.\label{onecoh}
\end{eqnarray}
The only novelty is that the fixed constant $\alpha_{10}$ (i.e. the failed clock) now acts as the inverse of the speed of propagation of the wave packet in the semi-classical time and space. 
This solution saturates, with equally spread uncertainties, the standard uncertainty principle in the variable with the coherent amplitude:
\be
\sigma_{T_2}^2=\sigma_{\alpha_2}^2 =  \frac{ \plk}{ 2 }.
\ee
All of this is unsurprising because having an infinitely sharp constant is the same as not having deconstantized that constant all,
so that its conjugate is a ``failed'', or uniformly distributed clock. Thus, we are left with a single clock and so we
recover the usual result~\cite{Jletter}. 

We could of course have chosen either of the two clocks as the failed clock. Note that the classical equations of motion can be written either as: 
\be\label{XT1}
\alpha_1 \dot X=T_2
\ee
or 
\be\label{XT2}
\alpha_2 \dot X= T_1
\ee
so both options reproduce the classical limit. Indeed the Hamilton equations for the system are:
\begin{eqnarray}
&& \dot{b}=\{b,H\}=-2N\frac{\alpha_1^2\alpha_2^2}{a^5} \label{eqb} \\
&& \dot{T_1}=\{T_1,H\}=-2N\frac{\alpha_1\alpha_2^2}{a^3} \label{eqT1}\\
&& \dot{T_2}=\{T_2,H\}=-2N\frac{\alpha_1^2\alpha_2}{a^3} \label{eqT2}, 
\end{eqnarray}
with the (Hamiltonian) constraint:
\be
b^2+k=\frac{\alpha_1^2\alpha_2^2}{a^4}.
\ee
Since, from the definitions of $X$ and $h(b)$:
\be
\dot X=\frac{\dot b}{\sqrt{b^2+k}}
\ee
we can easily prove that the motion of the peak of either coherent packet (i.e. Eq.~(\ref{XT1}) or (\ref{XT2})) represents
(\ref{eqb}), once we use the fact that for a coherent state (\ref{eqT1}) or  (\ref{eqT2}) have 
minimal uncertainty (beyond the obvious fact that they must be true on average, as per Ehrenfest's theorem~\cite{Jletter}).

\subsection{Two independent coherent clocks}
But what if we attempt to use the two semi-classical clocks concurrently?
Choosing a factorizable coherent state in the two constants:
\begin{eqnarray}
A(\alpha_1,\alpha_2)=\sqrt{\boldsymbol{N}(\alpha_{10},\sigma_1)}\sqrt{\boldsymbol{N}(\alpha_{20},\sigma_2)}
\end{eqnarray}
(so that $\sigma^2(\alpha_i)=\sigma^2_i$), and using
\begin{eqnarray}
\int d^nxe^{-\frac{1}{2}A_{ij}x^ix^j+B_ix^i}=\sqrt{\frac{(2\pi)^n}{\det(A)}}e^{\frac{1}{2}B^i(A^{-1})_{ij}B^j}
\end{eqnarray}
with
\begin{equation}
A_{ij}=
\begin{bmatrix}
\frac{1}{2\sigma_1^2} & \frac{-iX}{\plk} \\
\\
\frac{-iX}{\plk} & \frac{1}{2\sigma_2^2} \\
\end{bmatrix}
\end{equation}
and 
\begin{eqnarray}
B_i=\frac{\alpha_{i0}}{2\sigma_i^2}-\frac{i}{\plk}T_i
\end{eqnarray}
we can integrate (\ref{psi}) into:
\begin{widetext}
\begin{eqnarray}
&& \psi(b,T_1,T_2)=\sqrt{\frac{2\sigma_1\sigma_2}{\pi(\plk^2+4\sigma_1^2\sigma_2^2X^2)}}\exp\left[-\sum_{i=1}^{2}\frac{\alpha_{i0}^2}{4\sigma_i^2}\right]\cdot \notag \\
&& \cdot\exp\left[\frac{1}{2\det(A)}\left(\frac{1}{2\sigma_2^2}\left(\frac{\alpha_{10}}{2\sigma_1^2}-\frac{iT_1}{\plk}\right)^2+\frac{1}{2\sigma_1^2}\left(\frac{\alpha_{20}}{2\sigma_2^2}-\frac{iT_2}{\plk}\right)^2\right)\right] \cdot \notag \\
&& \cdot \exp\left[\frac{iX}{\plk\det(A)}\left(\frac{\alpha_{10}}{2\sigma_1^2}-\frac{iT_1}{\plk}\right)\left(\frac{\alpha_{20}}{2\sigma_2^2}-\frac{iT_2}{\plk}\right)\right],
\end{eqnarray}
\end{widetext}
with $\det(A)=\frac{1}{4\sigma_1^2\sigma_2^2}+\frac{X^2}{\plk^2}$.
This can be repackaged as
\begin{widetext}
\begin{eqnarray}\label{psitotfinal}
&& \psi(b,T_1,T_2)=\sqrt{\frac{2\sigma_1\sigma_2}{\pi(\plk^2+4\sigma_1^2\sigma_2^2X^2)}}
\exp\left[-\frac{\sigma_1^2(T_1-\alpha_{20}X)^2+\sigma_2^2(T_2-\alpha_{10}X)^2}{\plk^2+4\sigma_1^2\sigma_2^2X^2}\right]\cdot \notag \\
&& \cdot\exp\left[\frac{i\plk}{\plk^2+4\sigma_1^2\sigma_2^2X^2}\left(\alpha_{10}\alpha_{20}X-\alpha_{10}T_1-\alpha_{20}T_2-\frac{4\sigma_1^2\sigma_2^2}{\plk^2}T_1T_2X\right)\right]
\end{eqnarray}
\end{widetext}
with the interesting result that although the wave function does not factor, the cross terms are all in its phase (suggesting interesting entanglement effects), so that the probabilities do factor. 

Thus, we arrive at a modified saturated Heisenberg uncertainty identity:
\begin{eqnarray}
\sigma_{Ti}^2\sigma_{\alpha i}^2=\frac{\plk^2}{4} +\sigma_{\alpha 1}^2\sigma_{\alpha 2}^2X^2.
\end{eqnarray}
It reduces to the usual  one (i.e., Eq.(\ref{Heis0})) should one of the $\sigma_{\alpha i}$ vanish (i.e. one of the clocks be a failed clock), as expected\footnote{Note that a failed clock can be a coherent squeezed state: infinitely squeezed towards zero $\sigma_{\alpha}$.}.
However the situation is totally different when two clocks are at work:  then there  is an excess-uncertainty in the uncertainty relations of both clocks/constants due to their simultaneous use, which only vanishes when $X=0$.

For $X\gg \plk/(\sigma_1\sigma_2)$ the situation is very different. The peak of the distributions is still on the classical trajectory, but the probability becomes Gaussian in $T_i/X$, rather than in $T_i$ and $X$. The limiting form of the probability distribution is now:
\be
P(b,T_1,T_2)\propto \exp{\left[-\frac{{\left(\frac{T_1}{X}-\alpha_{20}\right)^2}}{2\sigma_2^2}\right]}
\exp{\left[-\frac{{\left(\frac{T_2}{X}-\alpha_{10}\right)^2}}{2\sigma_1^2}\right]}.\nn
\ee
The simplicity of this form cannot hide the fact that strong distortions from Gaussianity are now present in terms of the original measures $X$ and $T$. It is as if in this limit the splitting into space and time has to be done differently. Whether this is general or a peculiarity of this system is not clear to us.



\subsection{Other semiclassical states}\label{field-redef}
There are other semiclassical states but they all end up reducing the clocking system to a single clock. 
A notable example is the theory resulting from absorbing $G_N$ in a field redefinition:
\be
\tilde \phi =\sqrt G_N\phi,
\ee
so that it looks as if the $G_N$ disappears. Ignoring the issue of other fields, and of their possible interactions and self-interactions,
it should be obvious that such an operation leads to a different quantum theory, the states of which are a subspace of the states we started
from, the $A(G_N,\pi_\phi)$ forced to take the form $F(\pi_\phi/\sqrt{G_N})$, as we now explain in more detail.

Indeed, a field redefinition is an example of a general change of ``constant'' basis:
\be\label{betaofalpha}
{\bm \alpha}\rightarrow {\bm\beta}(\bm \alpha)
\ee
in the prescription:
\be
S_0\rightarrow S=S_0+ \int d^4 x\, {\bm \alpha} \cdot \partial_\mu {\bm T}^\mu,
\ee
where the function $\bm \beta$ can be degenerate. 
All theories resulting from such changes, albeit classically equivalent, lead to different quantum theories. 
Their measures and inner products are different, and states which are coherent in one basis will not be coherent in the other. 
Their spatial solutions, $\psi_s$, however, are still all the same.

Hence we can choose  a degenerate transformation reducing our two times to a  single one, such that its
conjugate matches the wave number of $\psi_s$ (as in (\ref{psis})), that is choose a single $\beta_1$:
\be
\beta_1=\alpha_1\alpha_2=\frac{\pi_\phi}{\sqrt{2\tilde G_N}}.
\ee
It is not difficult to see that the time, or momentum conjugate to $\beta_1$ is nothing but the redefined field $\tilde \phi$. 
These are natural constant and time variable of the system, since the general solutions then become:
\begin{eqnarray}
\psi(b,T_{\beta1})=\int d\beta_1 A(\beta_1)e^{\frac{i}{\plk}\beta_1 [X(b) - T_{\beta 1}]}.
\label{psi1}
\end{eqnarray}
The standard semi-classical results for one clock follow from the choice $A(\beta_1)=\sqrt{\boldsymbol{N}(\beta_{10},\sigma_1)}$.
The classical limit is reproduced by $\dot X(b) =\dot T_{\beta 1}$ in the usual way.

In fact we do not even need to make the transformation degenerate. We could choose any other independent $\beta_2$, and 
it would drop out of $\psi_s$. Then the second clock (conjugate to $\beta_2$) is a bad clock, not because it is failed (indeed we could choose a coherent state for it),  
but because it becomes redundant. Its wave function does not engage with $X$. The passing of its time is not associated with
any change in $X$, and the clock would simply factorize in 
\begin{eqnarray}
\psi(b,T_{\beta1},T_{\beta 2})=\int d\beta_1 d\beta_2 A(\beta_1,\beta_2)e^{\frac{i}{\plk}\beta_1 [X(b) - T_{\beta 1}]},
\label{psi12}
\end{eqnarray}
(assuming that $ A(\beta_1,\beta_2)$ factorizes).

\section{More general fluids}
\label{genfluids}
The situation described here for a deconstantized $G_N$ and a massless scalar field (for which $w=1$) can be generalized 
to a perfect fluid with any $w$. This can be seen in two ways, which although not equivalent, lead to the same result. 

For example we can stay close to the scalar field description and model such a fluid by 
$K$-essence model~\cite{k-ess}, in which we take a power of 
\be
X=-g^{\mu\nu}\partial_\mu\phi \partial_\nu\phi
\ee
as the kinetic term $K=K(X)$ in the action. Choosing:
\be
K=\frac{w }{1+w}X^{\frac{1+w}{2w}}
\ee
we end up with the $\phi$ action in mini-superspace:
\begin{eqnarray}
S&=&\frac{3V_c}{8\pi G_0}\int dt a^3\tilde G_NK
\nn\\
&=& \frac{3V_c}{8\pi G_0}\int dt\left[
\dot{\phi}\pi_\phi-Na\left(C_w \frac{\pi_\phi^{1+w}}{2 \tilde G_N^wa^{1+3w}}\right)\right] \nn
\end{eqnarray}
with
\begin{eqnarray}
&& \pi_\phi=\tilde{G}_Na^3 \left(\frac{\dot{\phi}}{N}\right)^{\frac{1}{w}}
\end{eqnarray}
(where $C_w$ is an irrelevant coefficient) instead of (\ref{piphi}). We would therefore obtain the usual solutions (\ref{psi1}), with $\beta_1=m$ where:
\be\label{mK}
m=\frac{\pi_\phi^{1+w}}{G_N^w}.
\ee
The only case in which a $G_N$ clock could not be used is that of a dust fluid, with $w=0$. 

We get the same result using the description employed in~\cite{gielen,gielen1}, itself taken from~\cite{brown}. Then the general action \cite{gielen1} :
\begin{widetext}
\be
S_{isent}=\frac{1}{G_0}\int d^4x\left[-G_N\sqrt{-g}\rho\left(\frac{|J|}{\sqrt{-g}}\right)+G_NJ^\mu(\partial_\mu\theta+\beta_A\partial_\mu\alpha^A)\right],
\label{sisent}
\ee
\end{widetext}
where $J^\mu$ is the densitised particle number flux and $\theta$, $\beta_A$ and $\alpha^A$ are Lagrange multipliers. $\rho$ is the energy density of the fluid and is a function of $|J|=\sqrt{-g_{\mu\nu}J^\mu J^\nu}$ and $\sqrt{-g}$.
In mini-superspace $J^\mu=a^3n\delta^\mu_t$ where $n=n(t)$ is the particle number density \cite{turok}. Thus, the action (\ref{sisent}) reduces to:
\bea
S_{isent}&=& \frac{3V_c}{8\pi G_0}\int dt\left[\tilde G_N a^3 n\dot \theta -Na^3\rho(n)\tilde G_N\right]
\eea
implying the conserved variable:
\be
C=\tilde G_Na^3 n
\ee
and we dropped the last term involving the other Lagrange multipliers.
For a perfect fluid with $p=w\rho$ we have $\rho=\rho_0n^{1+w}$, for some constant $\rho_0$. Hence,  the Hamiltonian can be written:
\be
H_{pf}=a^3\rho \tilde G_N=\frac{\rho_0}{a^{3w}\tilde G_N^w}
\ee
so that defining:
\be
m=\rho_0 C^{1+w}
\ee
we finally have, for some $\chi$:
\bea
S_{pf}&=& \frac{3V_c}{8\pi G_0}\int dt\left[\dot m  \chi -Na \frac{m}{\tilde G_N^w a^{1+3w}}\right]
\eea
in agreement with (\ref{mK}).

One may object to both actions on different grounds, but it is interesting that for our purposes it does not make any difference. 
As we see the discussion for a scalar field then generalizes for any fluid, with:
\be\label{GNw}
\alpha_1\propto G_N^{-w/2}
\ee
and $\alpha_2$ a suitable power of the fluid invariants variables, $\pi_\phi$ or $m$ depending on the description. 

\section{Sub-clocks and multicomponents with the same $w$}

\label{multicomp}

A final example is a Universe filled with different components with the same $w$, so that they all dominate at the same epoch. 
In principle, each one of them could be used to produce a separate quantum clock (all of them proportional to each other on-shell). A simple illustration is provided
by a set of $N$ massless non-interacting scalar fields, $\phi_i$, with conjugate momenta $\pi_i$ (keeping
$G_N=G_0$ fixed, for simplicity). The mini-superspace action has the form:
\begin{eqnarray}
S &=& \frac{3V_c}{8\pi G_0}\int dt\left[\dot{b}a^2+\sum_i\dot{\phi}_i\pi_i - H \right] \nn
\end{eqnarray}
where each $\phi_i$ can be seen as a relational clock, conjugate to ``constant'' $\pi_i$. 
The Hamiltonian constraint:
\begin{eqnarray}
H&=&Na\left( -(b^2+k)+\frac{1}{2a^4\tilde{G}_0}\sum_i\pi_i ^2\right)=0, 
\end{eqnarray}
can be placed in the form  $h(b)a^2-\alpha=0$ 
with the same $h(b)$ and  $X(b)$ used before (cf. Eqs.~(\ref{hofb}) and (\ref{Xofb})) and:
\be
\alpha= \sqrt{\sum_i\alpha_i^2}=\sqrt{\sum_i\frac{\pi_i^2}{2\tilde G_0}}.
\ee
Thus, the spatial wave function $\psi_s$ will be the usual plane wave in $X(b)$ for scalar fields, but  with a
``wave-number'' $\alpha$  which does not match any of the frequencies $\alpha_i$ associated with each partial clock. 

As a result,  the  most general wave-function (complete with time/frequency factors) is the superposition:
\begin{widetext}
\begin{eqnarray}
\psi(b,T_1,...,T_N)=\int d\alpha_1...d\alpha_NA(\alpha_1,...,\alpha_N)\exp{\left[i\frac{3V_c}{l_P^2}\left(X(b) \sqrt{\sum_i\alpha_i^2}-\sum_i\alpha_iT_i\right)\right]}
\end{eqnarray}
\end{widetext}
and the choice:
\begin{equation}
A(\alpha_1,...,\alpha_N)=\prod_i\frac{1}{(\sqrt{2\pi}\sigma_i)^{1/2}}\exp\left[-\frac{(\alpha_i-\alpha_{i0})^2}{4\sigma_i^2}\right]
\end{equation}
leads to a non-Gaussian state:
\begin{widetext}
\begin{eqnarray}
\psi(b,T_1,...,T_N)\propto\int d\alpha_1...d\alpha_N\exp\left[B\sqrt{\sum_i\alpha_i^2}-\sum_iC_i\alpha_i^2+\sum_iD_i\alpha_i\right],
\end{eqnarray}
\end{widetext}
where the coefficients $B$, $C_i$ and $D_i$ depend on $\alpha_{0i}$, $\sigma_i$, $T_i$ and $X(b)$, with the problems we have already highlighted.

To produce a coherent state we proceed as in Section~\ref{field-redef}, and  seek a canonical transformation 
into a time variable that has conjugate $\alpha$.
This could be obtained by writing the $\alpha_i$ in polar coordinates
and absorbing the radial part of the Jacobian into a redefinition of the amplitude. 
Then, all the times associated with the angles can either be discarded (with a degenerate transformation), or, if still present,
become irrelevant, since they factor out of the wave function, just as in Section~\ref{field-redef}. 
The wave function (or its relevant part that is not a phase factor) becomes:
\be
\psi(b,T_\alpha)=\int d\alpha A(\alpha)e^{\frac{i}{\plk}\alpha(X(b)-T_\alpha)}
\ee
and now a Gaussian $A(\alpha)$ does produce a coherent state. 
Using the standard expression for the transformation of the momenta for this canonical transformation~\cite{goldstein}:
\be
 {\bm T}^\mu_{\bm\beta}=\frac{\delta \bm\alpha}{\delta \bm\beta}{\bm T}^\mu_{\bm\alpha}
\ee
we find that the clock conjugate to $\alpha$ is:
\be
\Phi = \sum_i \frac{\alpha_i}{\alpha} \phi_i.
\ee

The situation we have illustrated for scalar fields applies to any fluid for which the linearizing wave-numbers $\alpha_i$ are not additive.
This is the general rule with the exception of radiation.  Indeed, as in~\cite{Jletter}:
\begin{eqnarray}
&& -(b^2+k)+\frac{1}{a^{1+3w}}\sum_im_i=0 \\
\implies
&& (b^2+k)^{\frac{2}{1+3w}} a^2-\left(\sum_im_i\right)^{\frac{2}{1+3w}}=0.
\end{eqnarray}
Radiation is exceptional in that the general wave function does factorize into 
wave functions for the different components:
\begin{widetext}
\begin{eqnarray}
\psi(b,T_1,...,T_N)=\int dm_1...dm_NA(m_1,...,m_N)\exp\left[\frac{i}{\plk}\sum_im_i(X_r(b)-T_i)\right]
\end{eqnarray}
\end{widetext}
so that a coherent state in each leads to
\begin{widetext}
\begin{eqnarray}
\psi(b,T_1,...,T_N)=\prod_i(8\pi\sigma_i^2)^{1/4}\exp\left[\frac{i}{\plk}(X_r-T_i)m_{i0}-\frac{\sigma_i^2}{\plk^2}(X_r-T_i)^2\right],
\end{eqnarray}
\end{widetext}
where $X_r=\int\frac{db}{b^2+k}$.
For any other $w$, {\it including the cosmological constant}, we have a problem using partial clocks, and we should use the full
clock matching all the components appearing in the Hamiltonian.

We finally note that a problem inevitably occurs if not all components with the same $w$ appearing in the Hamiltonian are clocks. 
Then, we can never find a frequency matching the  spatial wavenumber and (unless the $m$ are additive, as in the case of radiation) we have a problem.

\section{Relation between our coherent states and the WKB expansion}
In this paper we used the Heisenberg bound as a  
condition for classicality, as well as coherent states as a way to probe this bound. 
Clearly when a solution vastly exceeds this bound one cannot have a classical state. But how does this method
relate to the WKB criterium for classicality?

Historically, coherent states have rarely been used in standard quantum cosmology because,
with the most obvious definition (a state which is an eigenstate of the annihilation operators for each pair of non-commuting variables), they do not solve the Wheeler-DeWitt equation\fn{Less rigid definitions \cite{Kieferwp,Thiemanncoh} do exist and have been used.}. For example, for the pair $b$ and $a^2$, the eigenstates of $\hat z=\hat b+i \hat a^2$ (which are kinematically defined, and so do not have to satisfy the constraints) are not solutions of the Wheeler-DeWitt equation. The fact that such states are solutions for the unimodular-like pairs $\{\alpha,T_\alpha\}$ 
results from the fact that $T_\alpha$  does not appear anywhere in the dynamics. 

Hence we use coherent states in this context because we can, and without prejudice to the good work on 
assessing quantum vs semi-classical behaviour based on the WKB approximation (e.g.~\cite{KieferReview,Jonathan,Jonathan1})
in the context of standard GR.  But how does the WKB approach fare in the context of constant-time pairs?
We examine the example in Section~\ref{quant}, but what follows applies to any other unimodular-like situation. 
Let us consider the Schrodinger equation (\ref{doubleTScho}) and try out the WKB ansatz:
\be
\psi={\cal N} \exp{\left(\frac{1}{\plk}\sum_n\plk^n S_n
\right)}.
\ee
To order $n=0$ we have:
\be
-i \frac{\partial S_0}{\partial X}+  \frac{\partial S_0}{\partial T_1} 
\frac{\partial S_0}{\partial T_2} =0
\ee
so that:
\be
S_0= i [X(b)\alpha_{10}\alpha_{20}-\alpha_{10}T_1-\alpha_{20}T_2],
\ee
where $\alpha_1=\alpha_{10}$ and $\alpha_2=\alpha_{20}$ are fixed. Hence the monochromatic partial waves in (\ref{psi}) are 
WKB zeroth order solutions and these are exact.
This is unusual. It is due to the use of the connection 
representation. Even for a pure Lambda we know that the real Chern-Simons state is an exact sine wave~\cite{realCS}; the Airy function solutions 
obtained in the metric representation are not, and hence the need of a WKB approximation~\cite{Vilenkin}. From the point of view of the $b$ representation, the approximation is in the Fourier transform into $a^2$~\cite{CSHHV}. 

Now the problem is that the most general solution in a deconstantized setting (but not in standard GR) is a superposition of $n=0$ solutions 
and such superpositions must be considered to obtain a normalizable solution. So, even though $S_n=0$ for $n>0$ provides an unexpected exact solution (and this
is all there is in the standard GR, non-deconstantized setting), we should seek other solutions for $n\ge 1$. 
To order $n=1$ we find:
\be
-i \frac{\partial S_1}{\partial X} +\frac{\partial S_0}{\partial T_1} \frac{\partial S_1}{\partial T_2} +\frac{\partial S_0}{\partial T_2} \frac{\partial S_1}{\partial T_1} +\frac{\partial^2 S_0}{\partial T_1\partial T_2} =0
\ee
that is:
\be
\frac{\partial S_1}{\partial X} +\alpha_{10} \frac{\partial S_1}{\partial T_2}
+\alpha_{20} \frac{\partial S_1}{\partial T_1}=0.
\ee
Besides the trivial $S_1=0$, we can have any solution of the form:
\be\label{WKBS1}
S_1=F_1(\alpha_{10}X-T_2)F_2(\alpha_{20}X-T_1),
\ee
where $F_1$ and $F_2$ are general functions. 
Nothing forces these functions to be quadratic, and if quadratic nothing is said about squeezing. 
The only condition asked is that the amplitudes dressing the plane wave be {\it any} function peaked on the classical trajectory 
(that is,  Eqs.~(\ref{XT1}) and (\ref{XT2}), assuming the $T_i$ satisfies the classical equations of motion).
The WKB solution could be very non-classical indeed. 

This is to be contrasted with our exact solution (\ref{psitotfinal}), which is a special case of (\ref{WKBS1}) and so a WKB state too. 
Again, with this choice we have  that $S_n=0$ for $n\ge 2$ in the WKB expansion. Our treatment is therefore more exact and more discerning regarding the quantum uncertainities of the system. 

\section{Conclusions}
In this paper we examined the implications of using more than one cosmological clock in the same epoch, using a simple model based on a massless scalar field coupled to standard Einstein gravity, augmented by a ``deconstantized'' gravitational constant $G_N$ mimicking the prescription used in unimodular gravity to deconstantize Lambda. This leads to two distinct clocks, one the scalar field itself and the other the conjugate momentum to $G_N$. 

We then explored solutions to the double-time Schrodinger equation, with the view to drawing conclusions on the pros and cons of using multiple clocks simultaneously. Considering one constant to be infinitely sharp and the other a coherent state turned out to be trivial as this is equivalent to using just one clock (the other being a failed clock). Considering the product of two semi-classical states, instead, leads to a modified saturated Heisenberg uncertainty relation, dependent on the uncertainty of both constants and on the linearizing~\cite{DSR} spatial variable $X(b)$. Surprisingly, working with two clocks at the same time does not make our lives easier, since this increases the uncertainty associated with each clock"
working with a single clock is advisable. There are other well behaved semi-classical solutions, but they all reduce to a single effective clock, the remaining degrees of freedom corresponding to redundant clocks, the ticking of which does not correlate with changes in space. 

How general this might be is unclear, but we certainly were able to generalize our result in the most obvious directions. In particular, we explored a deconstantized $G_N$ faced with matter with other equations of state. We also considered Universes dominated by different fluids with the same equation of state, each supplying a clock. Again (with the notable exception of multi-radiation fluids) a single collective clock is preferable. The implications for the cosmological constant problem are under active investigation~\cite{weinberg,padilla}. The most obvious conclusion to draw would be the statement that quantum cosmology requires the existence of a total Lambda field, conjugate to a single total momentum, with obvious implications for the fine tuning of the partial Lambdas.

One might wonder whether clock independence would not be achievable by ``decoherence'' of the clocks.
But the analogy with decoherence is not quite applicable in the context of unimodular-like clocks, where classicality is 
achieved when the wave functions for the constants factorize and are suitably peaked around each "constant", but not so much so that the complementary clock/time cannot be also sharply peaked. The {\it relative} phase information between the two clocks is actually irrelevant for the matter.
Rather than decoherence, non-entanglement is what must be achieved for classicality and independence. 
A non-factorizable amplitude for the constants would lead to a very non-classical state. The different clocks, however,
still share the same ``space'' (the same $b$) which ultimately is why we obtained the results we did.

In closing, we note that the situation explored here differs from that in which different clocks are used at different epochs, ``one at a time'', as it were. As explained in~\cite{JPRD}, for the sake of a simple definition of unitarity and measure, one should in fact use different clocks at different epochs. What we seem to have shown here is that it is advisable to use one and only one clock in each epoch. There are of course situations where a clock seems handy throughout several epochs, for example the clock conjugate to $G_N$. But even then, we should only use it once, in one epoch. As the discussion in Section~\ref{genfluids} shows, different powers of $G_N$ would appear in the definition of coherent states in different epochs (cf. Eq.~\ref{GNw}). The state of $G_N$ could only be coherent in one of them.

{\it Acknowledgments.} We would like to thank  Steffen Gielen, Jonathan Halliwell and Tony Padilla for discussions in relation to this paper, as well
as an anonymous referee for urging us to write Section VII in the revised manuscript. This work was supported by FCT Grant 2021.05694.BD (BA) and STFC Consolidated Grant ST/L00044X/1 (JM).

\end{document}